\documentclass[12pt]{article}
\usepackage{fullpage}
\usepackage{epsfig}
\usepackage{amsmath}
\usepackage{amssymb}
\usepackage{color}

\begin{document}

\centerline{\bf\Large Simulations of light scalar mesons on the lattice}

\vspace{0.2cm}

\centerline{\bf\Large and related difficulties\footnote{Talk presented at {\it Scalar mesons and Related Topic} (SCADRON70) in Lisbon, Portugal, February 2008.}}

\vspace{0.5cm}

\centerline{\large Sasa Prelovsek}

\vspace{0.1cm}

\centerline{\small e-mail: {\it sasa.prelovsek@ijs.si}}

\vspace{0.1cm} 

\centerline{\small \it University of Ljubljana and Institute Jozef Stefan, Ljubljana, Slovenia}

\vspace{1cm} 

\centerline{\bf Abstract}

\vspace{0.3cm}

I review  the lattice simulations of light scalar mesons with $\bar qq$ and tetraquark interpolating fields. Several difficulties which complicate the extraction of scalar meson masses from the scalar correlators are pointed out. One of the major difficulties is the presence of the two-pseudoscalar scattering states, which often dominate the correlator at light quark masses. In the simulations with unphysical approximations such scattering contributions are even more disturbing since they are often large and sometimes negative. Techniques which allow extraction of scalar meson masses in presence of scattering states are listed. Preliminary  results of various lattice collaborations are presented.     

\section{Introduction}

\begin{figure}[htb!]
\includegraphics [height=5cm]{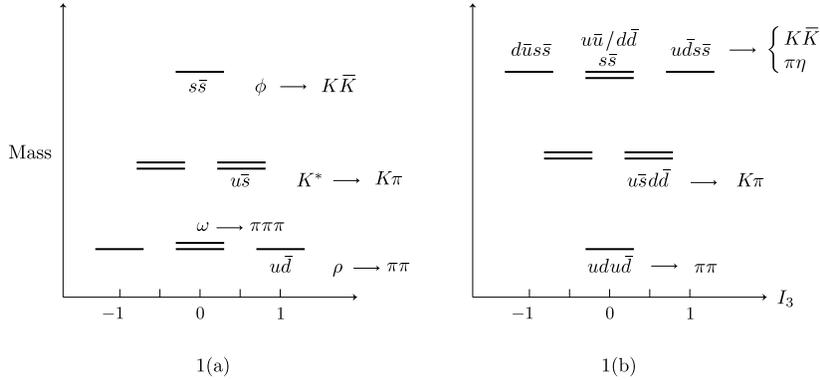}
\caption{ \small The illustration of the ``natural'' spectrum for (a) $\bar qq$ nonet, shown for the case of vector mesons, and (b) scalar tetraquark  nonet. } \label{fig1}
\end{figure}

The nature of light scalar mesons is not understood yet. Now there is a strong experimental evidence for a light $\sigma$ meson, whose pole was extracted with a small uncertainty from experimental data using model independent Roy analysis \cite{leutwyler}.  Experimentally  well established scalar resonances below $1$ GeV are also $a_0(980)$ and $f_0(980)$, while light $\kappa(800)$ is still controversial. At present it is still not clear whether these resonances represent the conventional $\bar qq$ nonet (its natural spectrum is illustrated in Fig. \ref{fig1}a for vector case) or perhaps the tetraquark states. The diquark anti-diquark tetraquarks $[qq]_{\bar 3_c,\bar 3_f}[\bar q\bar q]_{3_c,3_f}$ are theoretically especially well motivated: they lead to inverted mass spectrum illustrated in Fig. \ref{fig1}b, which has considerable similarity to observed spectrum \cite{jaffe,maiani}. The physical states may in reality be mixtures of $\bar qq$ and tetraquarks, while the singlets may mix also with glue. 

Lattice QCD has proven to be a very successful tool for determining the ground state hadron masses from first principles. The scalar meson puzzle would benefit from determination of scalar meson spectrum on the lattice, but I will argue that this is more challenging than determination of masses for most of other ground states. The basic question that lattice simulations raise in this respect is what is the mass of the lightest state that couples to 
a given $\bar qq$ or tetraquark interpolator. The major problem is that the lightest scalar state is usually a two-pseudoscalar scattering state. The information on the one-particle scalar resonance comes from sub-leading contribution and its extraction  is still challenging for present lattice simulations. 

In this paper I will briefly review results from the simulations of light scalar mesons on the lattice. Two nice reviews on this topic were  recently written by McNeile \cite{mcneile}.

\section{Methods for calculating scalar masses 
and related difficulties}

The basic object that is calculated on the lattice in order to extract the hadron mass is a hadron correlation function. Let me present the standard method for determining a hadron mass on the example of $a_0$ meson with $I=1$. This example will also point out the specific difficulties to determine the scalar meson masses due to the presence of scattering states. Correlation function  corresponds to creating a state $\bar d(\vec 0,0)u(\vec 0,0)$ at $t=0$ and annihilating it at some later time $t$
\begin{equation}
C(t)=\sum_{\vec x}\langle \bar d(\vec x,t)u(\vec x,t)|\bar u(\vec 0,0)d(\vec 0,0)\rangle~, 
\end{equation}
where total momentum $\vec 0$ is taken for concreteness. 
The interpolators with quantum numbers $J^{PC}=0^{++}$ and  $I=1$ may have various shapes as long as they couple with $a_0$.  For concreteness I have written the interpolators with point source and point sink. 
One of the main reasons why the scalar meson masses have not been yet reliably determined on the lattice is that the lattice correlators for light scalars and significantly more noisy than for light  pseudoscalars and vectors. The second major difficulty has to do with the presence of open or nearly open decay channels: 
the created state $\bar d(\vec 0,0)u(\vec 0,0)$ with $u$ and $d$ at the same point does not correspond to pure $a_0$ meson. It creates all eigenstates of the Hamiltonian  with quantum numbers $J^{PC}=0^{++}$. In simulations of proper QCD 
with $u,d,s$ dynamical quarks these are 
\begin{equation}
\label{eq1}
|\bar d(\vec 0,0)u(\vec 0,0)\rangle=A|a_0\rangle+B|a_0^*\rangle+C_i|\pi\eta\rangle_i+D_i|K\bar K\rangle_i+E_i|\pi\eta^\prime\rangle_i+...~,
\end{equation} 
where the  subscript $i$ refers to various momenta of pseudoscalars. The time evolution of the correlator is obtained by inserting the full set of Hamiltonian eigenstates $|n\rangle$ 
\begin{equation}
\label{eq2}
C(t)=\sum_n \langle |\bar du|n\rangle|^2~ e^{-E_nt}= |\langle \bar du|a_0\rangle|^2 e^{-m_{a0}t}+\sum_{i} |\langle \bar du|\pi\eta\rangle_i|^2 e^{-E_{\pi\eta}^i t}+ ...
\end{equation}
with $E_{\pi\eta}^{i}=m_\pi+m_\eta,\ \sqrt{m_\pi^2+\vec k_i^2}+\sqrt{m_\eta^2+\vec k_i^2}~,...$. Here $\vec k_{i}=2 \pi \vec j/L$ on the lattice with finite extent $L$. The scattering states with different momenta are well separated in energy for current lattice sizes.

If the lightest energy eigenstate was $a_0$, then the extraction of its mass 
 from $C(t)$ would be simple: one would just have to fit $m_{a0}$ from $C(t)\propto e^{-m_{a0}t}$ at large $t$, where states with larger energies have exponentially decreased. In nature, however, $m_\pi+m_\eta$ is smaller than $m_{a0}$ and the large-time behavior of the correlation function is dominated by the $\pi\eta$ scattering state. The lightness of two-pseudoscalar scattering states therefore  makes scalar meson spectroscopy particularly difficult. The scalar meson  mass has to be extracted from the sub-leading contribution to the correlator. The available methods to do that include:
\begin{itemize}
\item In the case of the scalar correlator with point source and point sink one 
can predict the two-pseudoscalar scattering state contribution using the corresponding version of Chiral Perturbation Theory (ChPT) up to a certain order of chiral expansion. The scalar meson mass can be determined by subtracting the 
predicted scattering contribution from the correlator and fitting that with $e^{-m_{a0}t}$.  This method has been first applied in quenched QCD by Bardeen et al. \cite{bardeen}. Two flavor and partially quenched QCD has been considered in \cite{sasaPQ} and three-flavor staggered QCD in \cite{sasaS,sasaMILC}. The scalar simulations with chiral valence quarks on staggered sea were studied in  \cite{sasaS,laiho,golterman}, while twisted-mass QCD in 
\cite{rehim,twisted}.  At light quark masses, the scattering states dominate the correlator and in all cases agree well with corresponding ChPT predictions. These are  most often  without any free-fit parameters, since they have all been pre-determined from other observables.   

\item  One- and two-particle states can be distinguished if correlator is computed at several lattice volumes since the spectral weight $|\langle\bar d u|n\rangle|^2$ of a two-particle state is proportional to $1/L^3$, while  
the spectral weight of one-particle state is not expected to be strongly volume dependent. Although the method is powerful, it has been only applied in \cite{liu} since results on various volumes are missing in dynamical simulations.  
\item 
Scattering states can also be identified by their energy shift as a function of lattice size and pseudoscalar mass. The method is costly and was only applied in \cite{jaffe,liu}.

\item One can apply different boundary conditions for quark and antiquark fields, which elevate the threshold for two-particle state with respect to one-particle state \cite{suganuma}.

\item A correlation matrix with various shapes of sources and sinks can be 
calculated instead of a single point-point correlator (\ref{eq2}). In this way the energy of the ground and some of the excited states can be extracted using the variational method, used for scalar spectra by UKQCD \cite{michael_a0,michael_a0_old,michael_singlet}, BGR \cite{bgr_quenched,bgr_dyn} and RBC \cite{hashimoto} collaborations.  
\end{itemize}
In the following I discuss the scalar meson simulations which use various techniques described above. 

\section{Simulations with $\mathbf{\bar qq}$ interpolators} 

{\bf $\mathbf{a_0}$ meson}

First I consider the {\bf simulations with $\mathbf{u,d,s}$ dynamical quarks} with relatively light $u,d$ quarks, where one hopes to be simulating nature as closely as possible. The only existing scalar meson simulations of this type use staggered valence quarks on a staggered sea (staggered QCD) \cite{sasaMILC} or chiral valence quarks on staggered sea quarks (mixed QCD) \cite{laiho}. The $m_{u,d}$ masses in these simulations are so low that one expects the $\pi\eta$ scattering state to have the lowest energy,  while the contribution of $a_0$  is expected to be sub-leading. Surprisingly, the 
ground state energies extracted from simulated correlators are not $m_\pi+m_\eta$, but closer  to $2 m_\pi$ for both types of simulations when $m_{u,d}$ are sufficiently low \cite{sasaMILC,laiho}. 
In addition, the correlator is negative 
in simulation that uses chiral quarks on staggered sea. 

Obviously, one is observing non-physical lattice artifacts, which get more and more pronounced as the $u,d$ masses approach thier experimental value. 
In order to understand what is the source of these unphysical effects, 
the contribution  of two-pseudoscalar intermediate state 
to $I=1$ point-point correlators at finite lattice spacing was predicted  
in corresponding  staggered and mixed ChPT  \cite{sasaS}. 
It was found that these artifacts are due to the discretization effects, 
which arise via taste violation and the use of different actions for valence 
and see quarks. In the continuum limit all $e^{-2m_\pi t}$ contributions were shown to 
cancel and analytic predictions agree with $\pi\eta$,$K\bar K$,$\pi\eta^\prime$ scattering given by conventional $ChPT$ \cite{sasaS}.  
At finite lattice spacing various $e^{-2m_\pi t}$ contributions do not cancel and analytical predictions  \cite{sasaS} agree well 
with simulated correlators in staggered \cite{sasaMILC} and mixed \cite{laiho} QCD. 
Let me point out that these analytical predictions, 
which largely dominate correlators at low $m_{u,d}$, 
have no free parameters at the lowest non-trivial order, 
since all of them have been determined 
by simulations of other quantities. The only free parameters of the fit 
are $m_{a0}$ and spectral weight $|\langle \bar d u|a_0\rangle |^2$. 
The extracted masses are $m_{a0}=1.01\pm 0.08$~GeV \cite{sasaS} and 
$m_{a0}=0.87\pm 0.15$ \cite{laiho}, 
but this can unfortunately not be  
very reliable values since the correlators are 
dominated by large and unphysical $e^{-2m_\pi t}$ contributions. 

\begin{table}[h]
\begin{tabular}{|c | c | c | }
\hline
Group & dyn. quarks & $m_{a0}$ \\
\hline
SCALAR Coll. \cite{scalar_col}  &  $u,d$ & $\simeq 2.4 m_\rho$   \\
Prelovsek et al. [RBC] \cite{sasaPQ} & $u,d$ & $1.58 \pm 0.34$ GeV \\
Prelovsek et al. [RBC] \cite{sasaPQ} & $u,d$ (part. quen.) & $1.51\pm 0.19$ GeV  \\
Hashimoto and Izubuchi [RBC] \cite{hashimoto} & $u,d$ & $1.111\pm 0.081$ GeV\\
McNeile, Michael, Hart 2003 [UKQCD] \cite{michael_a0_old} & $u,d$ & $1.0\pm 0.2$ GeV  \\
McNeile and Michael 2006 [UKQCD] \cite{michael_a0} & $u,d$ & $1.01\pm 0.04$ GeV \\
BGR 2007 \cite{bgr_dyn} & $u,d$ & $\simeq 1$ GeV \\
\hline
Bardeen et al. \cite{bardeen} & 0 & $1.34\pm 0.09$ GeV \\
BGR 2008 \cite{bgr_quenched} & 0  & $\simeq 1.4$ GeV\\
Mathur et al \cite{liu} & 0 & $1.42 \pm 0.13 $ GeV\\
\hline
\end{tabular}

\vspace{-0.1cm}

\caption{ \small Mass of the lightest $I=1$ scalar meson from simulations with dynamical $u,d$ quark and from quenched simulations.  }\label{tab1}
\end{table}

\vspace{0.1cm}

Even if the unphysical artifacts were not present, one expects that 
the correlators in simulations with $u,d,s$ dynamical quarks would be 
dominated by $e^{-(m_\pi+m_\eta)t}$ at light $m_{u,d}$. 
For this reason I would propose that 
{\bf simulations with only $\mathbf{ u,d}$ dynamical 
quarks} have more chances to determine $m_{a0}$, although they are 
less physical due to absence of dynamical $s$ quarks. The only 
two-pseudoscalar intermediate state in this case is $\pi\eta^\prime$, 
which is relatively heavy due to the anomaly and not so disturbing. 
The resulting masses from this type of simulations are shown in Table 
\ref{tab1}. The key question is of course what is the mass of the lightest 
state that couples to $\bar d u$: does it correspond to  
$a_0(980)$ or $a_0(1450)$?
The first three simulations \cite{sasaPQ,scalar_col} 
used a single correlator and were forced to fit at relatively low times, 
therefore the resulting mass might be artificially high due to the 
contribution of excited states. The rest of the two-flavor simulations 
\cite{michael_a0,michael_a0_old,bgr_dyn,hashimoto} calculated a correlation 
matrix and determined $m_{a0}$ using variational analysis, obtaining $m_{a0}\simeq 1$ GeV. The recent result from domain-wall fermions \cite{hashimoto} 
is un 
update of \cite{sasaPQ} with larger statistics and  $2\times2$ correlation
 matrix (using smeared and point source and sink): the resulting mass is lower, demonstrating the excited state contribution in the point-point correlator of \cite{sasaPQ}. McNeile and Michael \cite{michael_a0} determined $m_{b1}-m_{a0}=221(40)$ MeV on the lattice, compared to experimental result $245$ MeV, and  deduced $m_{a0}$ from experimental value of $m_{b1}$. Although it appears that the more recent lattice simulations get a mass close to $a_0(980)$, 
the question whether $a_0(980)$ couples to $\bar qq$ is still open. 
The energies in the Table \ref{tab1} 
simply correspond to the ground state energy, 
which may be $a_0$ or $\pi\eta\prime$ - both have comparable energy. 
Before final conclusion on $m_{a0}$ may be drawn, the two flavor 
simulations have to identify $a_0$ as well as $\pi\eta^\prime$. An attempt to 
estimate contribution of $\pi\eta^\prime$ to the point-point correlator was made in \cite{sasaPQ}: the ChPT renders the $\pi\eta^\prime$ contribution to be one order or more smaller than the lattice point-point correlators for  $m^{Nf=3}_0\ge 600$ MeV in the fitted time range.  

\vspace{0.1cm}

One would naively think that the {\bf quenched} correlator would be free of scattering states due to the lack of the sea quarks, but in this case the situation is even worse. The $e^{-2m_\pi t}$ contributions, which are large and negative, arise from unphysical $\pi\eta^\prime$ scattering in quenched QCD \cite{bardeen}. In Table \ref{tab1} I collect some of the recent quenched results for $m_{a0}$ which take these unphysical effects into account. A one-particle $a_0$ state was found in addition to two ghost scattering states in a variational analysis \cite{bgr_ghosts} of a $8\times 8$ scalar correlator matrix in \cite{bgr_quenched}. Analysis of  a single point-point overlap correlator allowed the authors of \cite{liu} to extract  two ghost states and a physical $a_0$ using a sequential empirical Bayes method. The quenched results for $m_{a0}$ are higher than recent dynamical results. 
 
\vspace{0.3cm}

{\bf $\mathbf{\kappa}$ and $\mathbf{K_0^*}$ mesons}

The relevant $\bar u s$ correlator receives contribution from strange scalar meson  as well as $\pi K$ scattering. The $\pi K$ state is so light that it has to be disentangled in any kind of simulation before  conclusion can be drawn about $m_\kappa$. So far the only simulation which demonstrated one and two particle states in  $\bar u s$ correlator was a quenched\footnote{Quenched $I=1/2$ correlator receives $e^{-(m_\pi+m_K)t}$  due to the unphysical $K\eta^\prime$ scattering state.} overlap simulation \cite{liu}, giving $m_{K_0^*}=1.41\pm 0.12$ GeV. 

\vspace{0.3cm}

{\bf $\mathbf{\sigma}$ meson}

Calculating the mass of the lightest singlet $\bar qq$ meson
 on the lattice is 
even more challenging than for the nonsinglet mesons, described above. 
The $\pi\pi$ scattering state is particularly light and has to be 
disentangled.  The correlator is a sum of connected quark diagram, which 
is very sensitive to any non-physical approximations undertaken in simulations, and the disconnected diagram, which is very expensive to evaluate numerically. On top of that the singlets mix with glueballs and their masses have  shown sizable lattice spacing dependence \cite{michael_singlet}.

A large scale simulation  of a single $\bar qq$ correlator with $u,d$ 
dynamical quarks on a fairly small lattice was performed by the 
SCALAR collaboration, giving $m_\sigma\simeq m_\rho$ with a single exponential fit \cite{scalar_col}. 
The extracted $m_\sigma$ is (slightly) below their $\pi\pi$ threshold. 
In order to be confident that observed ground state corresponds to $\sigma$, 
a further study is needed which would identify the $\pi\pi$ scattering state.

Lee and Weingarten \cite{weingarten} 
studied the mixing of $\bar qq$ and glue in a quenched 
simulation with quark masses around strange quark mass. They found that 
$f_0(1710)$ meson was $74(10)\%$ glueball. More recently a
dynamical simulation of $\bar qq$ and glue mixing was performed by 
UKQCD \cite{michael_a0} using the  variational analysis. 
They find the mass of approximately $1$ GeV for  
the lightest isosinglet state, which couples to both $q\bar q$ and gluonic 
interpolators. In comparison, the quenched simulations render glueball 
masses well above $1$ GeV. 

The singlet $\bar qq$ point-point correlator was simulated also with staggered valence quarks on a staggered sea \cite{sasaMILC}. 
As expected, the lattice correlator is largely dominated by $\pi\pi$ 
scattering  and agrees well with the $\pi\pi$ scattering prediction from staggered ChPT \cite{sasaMILC}.  The only unconstrained parameters in the fit are related to sub-leading $\sigma$ contribution, leading to $m_\sigma\approx 0.75\pm 0.16$ GeV. The  resulting $m_\sigma$ can unfortunately not very reliable due to the dominance of the $\pi\pi$ scattering, whose contribution is even more complicated due to the taste breaking at finite lattice spacing. 

ETM collaboration simulated $\bar qq$ meson with two flavors of dynamical 
twisted-mass quarks and found a mass close to $2m_\pi$ \cite{twisted}, 
which is  probably due to the $\pi\pi$ scattering.

\section{Simulations with tetraquark interpolators}

 So far all lattice simulations  
with tetraquark interpolators have been quenched. They also all omitted the 
contribution of the disconnected quark diagrams to the correlator. These two 
approximation can be even justified in this case to some extent: 
one can unambiguously assign a quark content to a state by excluding processes
 that mix $\bar qq$ and $\bar q^2q^2$ - dynamical quark loops and 
disconnected quark diagrams do exactly that. 

\begin{figure}[htb!]
\includegraphics [height=5cm]{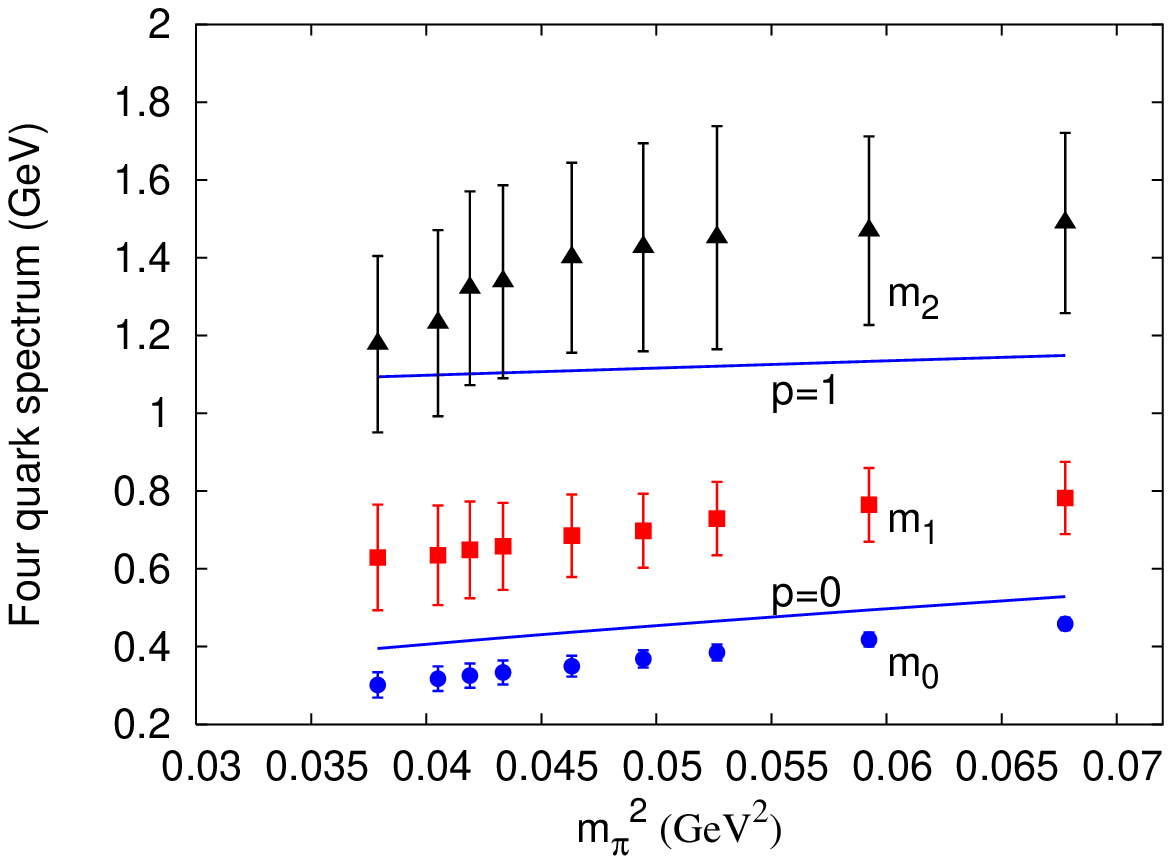}
\includegraphics [height=5cm]{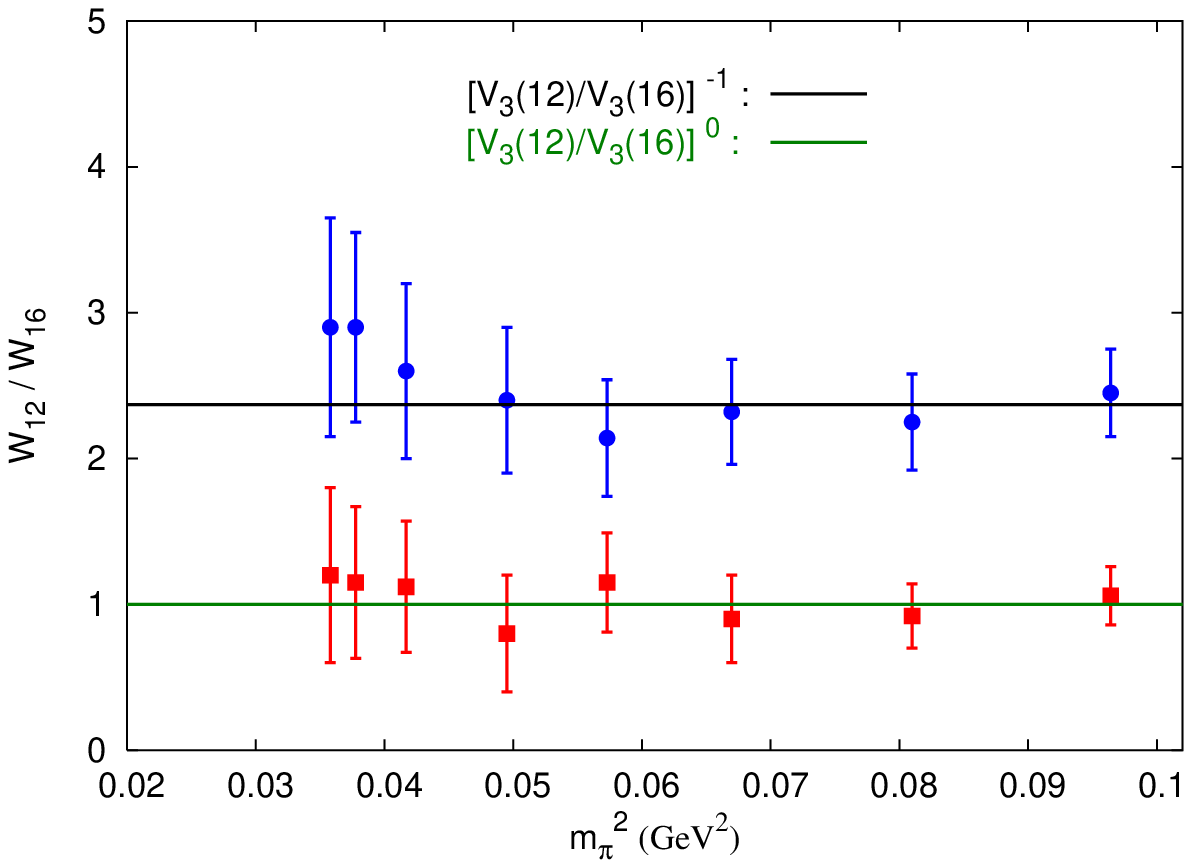}
\caption{ \small Three lowest energy states extracted from a single quenched tetraquark correlator \cite{liu} are shown on the left. The ratio of the spectral weights for two volumes in case of lower two states are shown on the right. } \label{fig2}
\end{figure}

Mathur and collaborators \cite{liu} calculated the $I=0$ correlator with interpolator $(\bar q\gamma_5q)(\bar q\gamma_5q)$ at the source and sink. They used overlap quarks with small $m_{u,d}$, corresponding to $m_\pi$ down to $182$ MeV. They analyzed correlators at spatial volumes $12^3$ and $16^3$.  At each volume, they were able to extract three states from a single point-point correlator using the sequential empirical Bayes method. The ground state had energy slightly lower than $2m_\pi$, while the ratio of the spectral weights for two volumes was consistent with $16^3/12^3$  (see Fig. \ref{fig2}). This indicates a $\pi\pi$ scattering state with zero relative momentum. The energy shift of this scattering state from the value $2m_\pi$ agrees with the analytical prediction from  quenched ChPT 
\cite{shift_quenched} at both volumes. 
The third state had an energy close to the energy of two pions with lowest non-zero momentum $2 \pi/L$. In the middle they found a state with energy well separated from any $\pi\pi$ scattering state and with a ratio of the spectral weights for two volumes consistent with one. This gives an indication of a one-particle tetraquark state with a mass  of about $600$ MeV, close to observed $\sigma$ \cite{leutwyler}. 
This is one of the few available lattice studies which was able to separate the one and two particle contributions to the correlator. 

Alford and Jaffe \cite{jaffe} simulated the correlator with the same  
$(\bar q\gamma_5q)(\bar q\gamma_5q)$ interpolator in non-exotic $I=0$ and 
exotic $I=2$ channels. The correlator was computed at a single relatively 
heavy quark mass and for a range of lattice sizes $L$. They studied
 the energy shift $\delta E=E-2m_\pi$  as a function of $L$, where $E$ is the ground state energy extracted from the lattice correlator. They found that 
$\delta E(L)$ for the exotic channel agrees with the full\footnote{It is surprising that the energy shift agrees with full ChPT prediction as one would have to compare it with quenched ChPT prediction. I guess this is due to the sizable mass of the quarks in \cite{jaffe}.} 
ChPT prediction 
for $\pi\pi$ scattering at finite volume. This  indicates that 
the $I=2$ ground state observed on the lattice is $\pi\pi$ scattering. 
The energy shift for $I=0$ does not follow the full ChPT 
analytic prediction for $\pi\pi$ scattering, providing another 
indication for the existence of a $I=0$ tetraquark.   

 \vspace{0.2cm}   

The Japanese group \cite{suganuma} made a simulation with  
diquark anti-diquark 
interpolator, which is very well theoretically motivated \cite{jaffe,maiani}. 
The  $I=0$ state in 
 quark mass range $m_s<m_q<2m_s$ is simulated. They used conventional  
boundary conditions (periodic for $q$ and $\bar q$) as well as 
hybrid boundary conditions (anti-periodic for $q$ and periodic for $\bar q$). 
The second option does not alter tetraquark energy while it 
raises two-pion threshold (both pions have nonzero momenta) \cite{suganuma}. 
They extracted a ground state energy from a single correlator and it was consistent with $\pi\pi$ scattering in both cases. This indicates that  a tetraquark resonance lies above the $\pi\pi$ threshold, if it exists in the quark mass region $m_s<m_q<2m_s$. 

 \section{Conclusions}

\vspace{-0.2cm}

Lattice QCD has not yet given a definite answer what are 
the masses of the lightest $\bar qq$ and tetraquark resonances 
with $J^P=0^{++}$ and  $I=0,1/2,1$. 
The major reason why the consensus has not been reached is that two-pseudoscalar scattering states usually represent the state with the lowest energy in the simulations. The information about the resonant states has to be extracted from sub-leading contributions to the correlators, which is still challenging for most of the current lattice simulations. A quenched lattice simulation which was able to get an indication for a tetraquark $\sigma$ resonance in addition to $\pi\pi$ scattering state  was presented in \cite{liu} 
(see Fig. \ref{fig2}). 
The unquenched simulations of correlation functions with a variety of sources and sinks on a range of volumes would allow separation of scattering and resonant states, therefore allowing determination of spectra for ground state scalar resonances. I believe this is a feasible project for the near future.    
  
 \vspace{1cm}

{\bf  \large Acknowledgments}

\vspace{0.3cm} 

  I would kindly like to thank  C. DeTar, C. Bernard, C. Gattringer and  
D. Mohler for reading the manuscript. 
  In addition, I would like to thank  C. DeTar, C. Bernard, Z. Fu, C. Gattringer, C. Lang, D. Mohler, L. Glozman, K. Orginos, C. Dawson, T. Izubuchi and A. Soni, with whom I had a pleasure to work on the scalar mesons during the past years.

\end{document}